\documentclass[reprint, amsmath, amssymb, aps, superscriptaddress, floatfix]{revtex4-2}%
\usepackage[dvipdfmx]{graphicx}
\usepackage{bmpsize}
\usepackage{color}
\usepackage{svg}
\usepackage{dcolumn}
\usepackage{bm}
\usepackage{amsmath}
\usepackage{amsfonts}
\usepackage{amssymb}%
\usepackage{braket}
\usepackage[caption=false]{subfig}
\usepackage{booktabs}
\usepackage{longtable}

\providecommand{\U}[1]{\protect\rule{.1in}{.1in}}
\begin{document}

\preprint{APS/123-QED}%
\title{Parent Hamiltonian as a benchmark problem for variational quantum eigensolvers}%

\author{Fumiyoshi Kobayashi}%
\affiliation{Graduate School of Engineering Science, Osaka University,
	1-3 Machikaneyama, Toyonaka, Osaka 560-8531, Japan.}%
\email{fkobayashi@qi.mp.es.osaka-u.ac.jp}%
\author{Kosuke Mitarai}
\affiliation{Graduate School of Engineering Science, Osaka University,
	1-3 Machikaneyama, Toyonaka, Osaka 560-8531, Japan.}
\affiliation{Center for Quantum Information and Quantum Biology,
	Institute for Open and Transdisciplinary Research Initiatives, Osaka University, Japan.}
\affiliation{JST, PRESTO, 4-1-8 Honcho, Kawaguchi, Saitama 332-0012, Japan.}
\author{Keisuke Fujii}
\affiliation{Graduate School of Engineering Science, Osaka University,
	1-3 Machikaneyama, Toyonaka, Osaka 560-8531, Japan.}
\affiliation{Center for Quantum Information and Quantum Biology,
	Institute for Open and Transdisciplinary Research Initiatives, Osaka University, Japan.}
\affiliation{Center for Emergent Matter Science, RIKEN, Wako Saitama 351-0198, Japan.}

\begin{abstract}%

Variational quantum eigensolver (VQE), which attracts attention as a promising application of noisy intermediate-scale quantum devices, finds a ground state of a given Hamiltonian by variationally optimizing the parameters of quantum circuits called ansatz. Since the difficulty of the optimization depends on the complexity of the problem Hamiltonian and the structure of the ansatz, it has been difficult to analyze the performance of optimizers for the VQE systematically. To resolve this problem, we propose a technique to construct a benchmark problem
whose ground state is guaranteed to be achievable with a given ansatz by using
the idea of parent Hamiltonian of a low-depth parameterized quantum circuits.
We compare the convergence of several optimizers by varying the distance of the initial parameters from the solution and find that the converged energies showed a threshold-like behavior depending on the distance.
This work provides a systematic way to analyze optimizers for VQE and contribute to the design of ansatz and its initial parameters.

\end{abstract}%

\maketitle

\section{\label{sec:Intro} Introduction}
In recent years, quantum computers are gaining attention as a hardware for next-generation information processing beyond classical ones. 
While quantum computing devices have made rapid progress in the last few years, 
their scale is still too small for fault-tolerant quantum computing.
Such current quantum computers are called noisy intermediate-scale quantum (NISQ) devices \cite{Preskill2018quantumcomputingin}.
NISQ-aware quantum algorithms are actively researched to seek practical applications of NISQ devices~\cite{Cerezo2021-ac}.

Variational quantum eigensolver (VQE) \cite{Peruzzo2014}, which is a variational algorithm to calculate an approximate ground state of a target Hamiltonian, is a promising candidate for NISQ applications.
In the VQE, parameters of a parameterized quantum circuit, frequently called ansatz, are optimized to minimize the expectation value of a Hamiltonian.
Therefore, the performance of the optimizer has a great impact on the performance of the VQE.
There are several proposals of optimizers specifically designed for the VQE \cite{Kubler2020adaptiveoptimizer, PhysRevResearch.2.043246, Nakanishi2019, koczor2020quantum, arrasmith2020operator, McArdle_2019, yamamoto2019natural, Stokes2020quantumnatural,Gu2021}.
However, those optimizers are frequently compared on a case-by-case basis.
To analyze the performance of these optimizers in a comparable way, we need systematic benchmark tasks.

The performance of optimizers in VQE depends mainly on two factors: the target Hamiltonian and the ansatz. For example, using an Ising model Hamiltonians that encode NP-hard problems makes the optimization NP-hard even for simple ansatz with only single-qubit rotations~\cite{Bittel2021}. Deep circuit tends to have more expressive power \cite{kim2020universal} but the optimization becomes difficult due to the notorious barren plateau problem \cite{mcclean2018barren}.
Therefore, for a systematic benchmark of the VQE optimizers, we need to construct a technique that can separate the above two factors.

To this end, we propose a method to generate benchmark tasks whose difficulty depends solely on given ansatzes.
In the proposed method, we generate Hamiltonians whose exact ground state is guaranteed to be achievable with a given parameterized quantum circuit.
Therenby, the proposed benchmark can purely test if a given optimizer can find the solution that is guaranteed to exist.
Specifically, we consider one-dimensional parameterized quantum circuits and calculate the parent Hamiltonian of its corresponding matrix product state (MPS) \cite{Perez-Garcia2007, ruiz2011tensor} with a given random parameters.
Importantly, unlike existing fidelity benchmarks \cite{Nakanishi2019, Kubler2020adaptiveoptimizer}, the problem can be constructed in a natural setting where VQE is usually performed.
We perform numerical experiments to demonstrate the validity and capability of the proposed method.
Specifically, we construct 7-local 12-qubit Hamiltonian for translation invariant parameterized quantum circuits with 6 independent parameters to compare the convergence of optimizers with varying the distance of the initial parameters from the solution.
We find that the converged energies show a threshold-like behavior; there is a threshold value for the distance from the solution, and if the solution is within that range, the optimal solution is reached, otherwise a sub-optimal solution is reached.
The proposed method can be straightforwardly applied for various low-depth ansatzes.
It would provide an essential knowledge 
about the convergence of variational quantum algorithms and would be helpful to improve the design of ansatzes and optimizers.

\section{\label{sec:method} Parent Hamiltonian-based benchmark}

\subsection{Variational quantum eigensolver}

The VQE is an algorithm for obtaining ground state energy on a quantum computer \cite{Aspuru-Guzik2005, Peruzzo2014, Kandala2017}.
Let an $N$-qubit parametric quantum circuit, called ansatz, be $U(\vec{\theta})$ where $\vec{\theta}$ is a set of parameters. 
The goal of VQE is to approximate ground state by $\ket{\psi(\vec{\theta})} = U(\vec{\theta})\ket{0}^{\otimes N}$.
VQE achieves this in the following way. 
For a given Hamiltonian $H$, an expectation value of energy $E(\vec{\theta}) = \bra{\psi(\vec{\theta})} H \ket{\psi(\vec{\theta})}$ is measured as a cost function for optimization. 
Then, parameters $\vec{\theta}$ are optimized to minimize $E(\vec{\theta})$ using a classical optimizer such as gradient-based methods \cite{Flet87} or non-gradient-based methods \cite{conn2009introduction}. 
For the gradient-based methods, the gradient of the cost function can be calculated by the so-called parameter shift rule \cite{Mitarai2018, Schuld2019, Mitarai2019}.

\subsection{\label{sec:method:MPS_parentHamiltonian}Matrix product state and parent Hamiltonian}
For clarity, we restrict our attention into one-dimensional parameterized quantum circuits and corresponding matrix product states. However, it is straightforward to extend the following argument for general low-depth quantum circuits.
An MPS of a system consisting of $N$ qubits with periodic boundary condition is defined as,
\begin{eqnarray}
\ket{\psi^{(N)}} = \sum_{i_1\cdots i_N= 0}^{1} \mathrm{Tr}[A^{[1]}_{i_1}A^{[2]}_{i_2}\cdots A^{[N]}_{i_N}]\ket{i_1, i_2, \cdots, i_N}
\end{eqnarray}
where $A^{[k]}_{i_k} \hspace{4pt} (0\le k\le N$) is a $D \times D$ matrix with $D$ being bond dimension.
A quantum state generated from a quantum circuit can generally be represented by an MPS, albeit its bond dimension can increase exponentially with respect to the circuit depth.

A Hamiltonian which has a given MPS as its ground state is called parent Hamiltonian of the MPS \cite{Perez-Garcia2007, ruiz2011tensor}. Let $\rho^{(n)}_i$ be a reduced density matrix of an MPS $\ket{\psi^{(N)}}$ on the $i$ th to the $\{(i+n) \mod N\}$ th qubits and $h_i$ be the projectors for $\mathrm{Ker}(\rho^{(n)}_i)$ of the reduced density matrix $\rho^{(n)}_i$ i.e. $h_i\ket{\psi^{(N)}} = 0$.
The parent Hamiltonian $H_{\mathrm{parent}}$ can be constructed as,
\begin{eqnarray}\label{eq:parent}
H_{\mathrm{parent}} \equiv \sum_{i=0}^{N-1} h_i.
\end{eqnarray}
$H_{\mathrm{parent}}$ has the MPS $\ket{\psi^{(N)}}$ as a ground state, and its energy is exactly zero.

\subsection{Recipe for the benchmark problem}
Below we explain how to construct benchmark problem for a given ansatz $\ket{\psi(\vec{\theta})}=U(\vec{\theta})\ket{0}$ using the parent Hamiltonian.
The idea is to generate a parent Hamiltonian by relating $\ket{\psi(\vec{\theta})}$ with an MPS.
The concrete procedure to construct the benchmark problem is as follows:
\begin{enumerate}
	\item Choose a set of answer parameters $\vec{\theta}_{\mathrm{ans}}$.\label{answer}
	\item Calculate $\mathrm{Ker}\,\rho^{(n)}_i$ of $\ket{\psi(\Vec{\theta})}$ for all $i$ starting from $n=1$ to sufficiently large $n$ until $\mathrm{Ker}\,\rho^{(n)}_i$ becomes non-null for all $i$.
	\item Search an orthnormal basis of $\mathrm{Ker}\,\rho^{(n)}_i$ and construct a parent Hamiltonian $H_{\mathrm{parent}}$ by Eq. \eqref{eq:parent}.

\end{enumerate}

By its very construction, the quantum circuit $U(\vec{\theta})$ can represent the exact ground state of $H_{\mathrm{parent}}$ generated by the above procedure by setting $\vec{\theta} = \vec{\theta}_{\mathrm{ans}}$.
While we considered a one-dimensional system that can be described by an MPS in the above, 
the benchmark problem with a guaranteed optimal solution can be constructed efficiently from parent Hamiltonian obtained from kernels of reduced density operators for low-depth parameterized quantum circuits with an arbitrary qubit connectivity.

It might be thought that the parent Hamiltonian obtained in this manner becomes merely a transformation of $-\sum_i Z_i$ under Heisenberg picture $\tilde{H} = U(\vec{\theta}_{\mathrm{ans}})\left(-\sum_i Z_i\right)U^{\dagger}(\vec{\theta}_{\mathrm{ans}})$.
If this is the case, the cost function becomes 
$$E(\vec{\theta}) = \bra{0}U^\dagger(\vec{\theta})U(\vec{\theta}_{\mathrm{ans}})\left(-\sum_i Z_i\right)U^{\dagger}(\vec{\theta}_{\mathrm{ans}})U(\vec{\theta})\ket{0},$$
which is almost equivalent to defining $E(\vec{\theta}) = \left|\bra{0}U^{\dagger}(\vec{\theta}_{\mathrm{ans}})U(\vec{\theta})\ket{0}\right|^2$, where they both check if $U^\dagger(\vec{\theta}_{\mathrm{ans}})U(\vec{\theta})\ket{0}$ returns $\ket{0}$ or not.
This ``trivial'' benchmark problem has been used in e.g. Ref. \cite{Nakanishi2019, Kubler2020adaptiveoptimizer}.
In numerical simulations presented in the following section, we show that this is not the case by checking locality of the resulting Hamiltonian.

A possible limitation of this benchmark is that the problem Hamiltonian, $H_{\mathrm{parent}}$, does not correspond to practical problems, and hence an optimizer that performs well on this benchmark may not perform equally well on such problems.
On the other hand, the previous researches of VQE optimizers \cite{Kubler2020adaptiveoptimizer, PhysRevResearch.2.043246, Nakanishi2019, koczor2020quantum, arrasmith2020operator, McArdle_2019, yamamoto2019natural, Stokes2020quantumnatural,Gu2021} have emparically shown that an optimizer that performs well on a specific problem performs also well for other problems.
We therefore believe that our proposal is suitable as a benchmark of VQE optimizers to a certain extent.

\section{\label{sec:numerical} numerical simulation}
\begin{figure}
\subfloat[\label{fig:blickwork}]{%
  \includegraphics[width=0.55\linewidth]{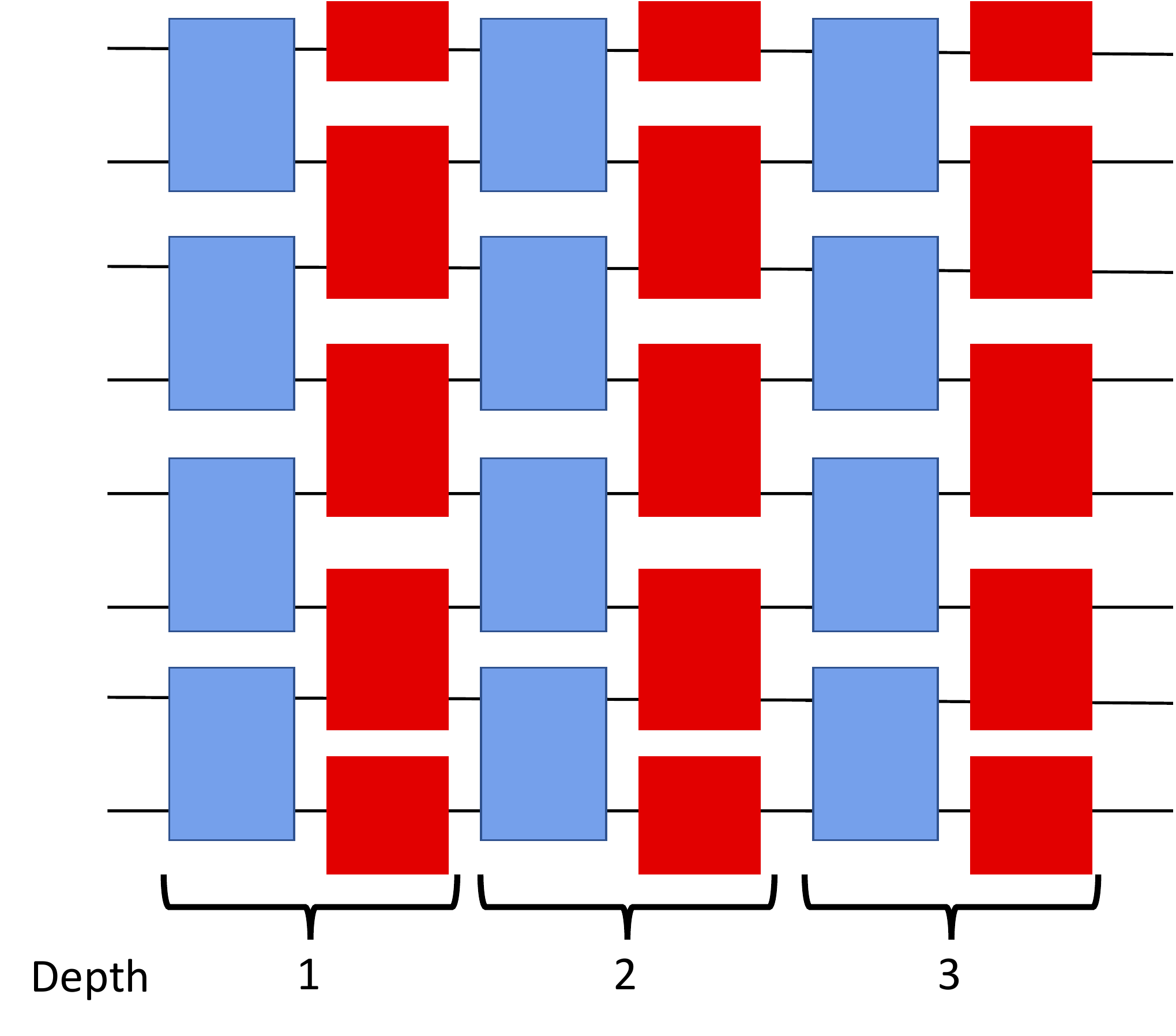}%
}\hfill
\subfloat[\label{subfig:b}]{%
  \includegraphics[width=0.4\linewidth]{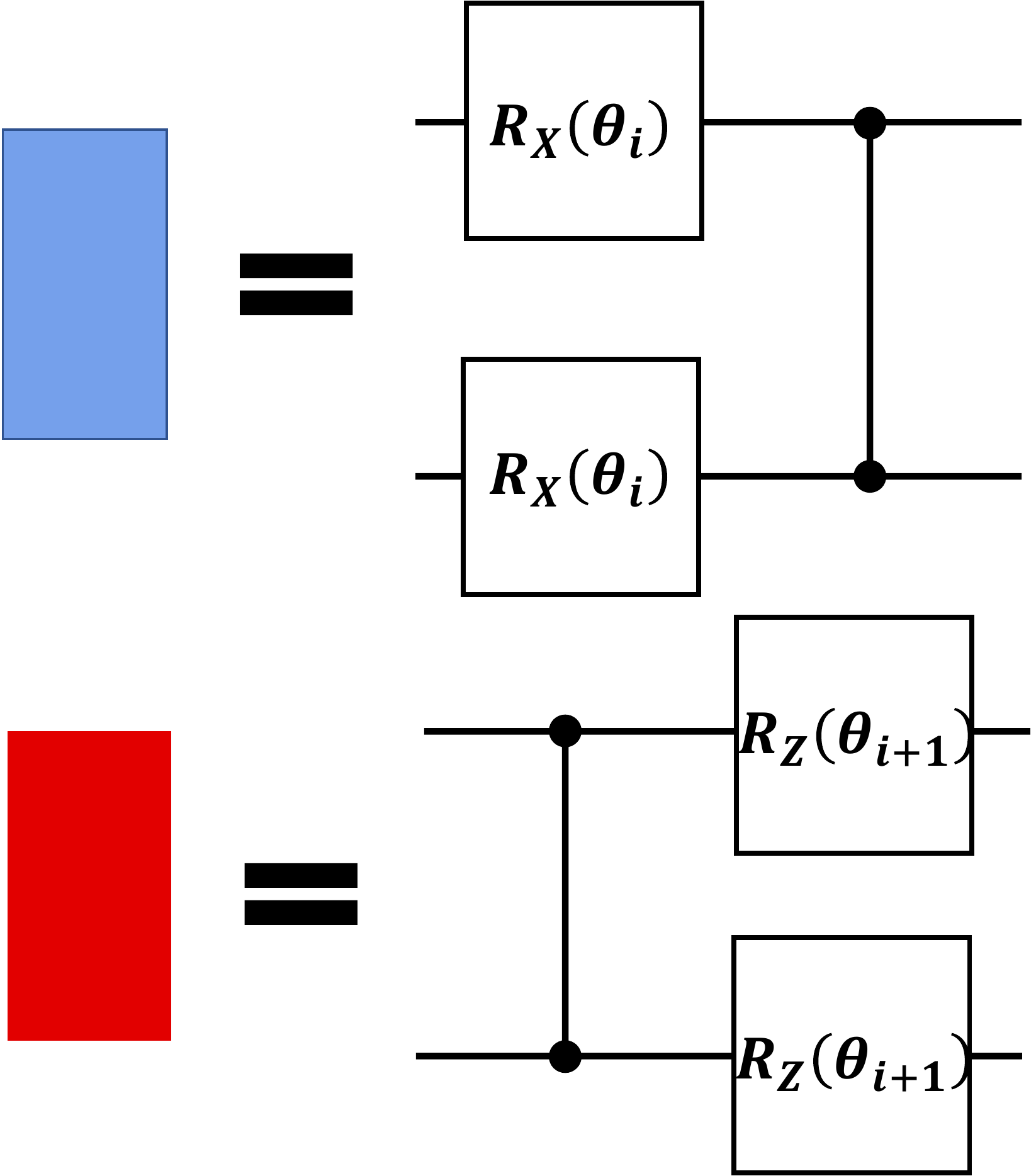}%
}
\caption{(a) The ansatz circuit $U(\vec{\theta})$ used in the numerical simulations. 
The top and bottom qubits share a 2-qubits gate in red ~(dark) blocks.
(b) The concrete construction of red~(dark) and blue~(light) blocks in (a). $R_X(\theta)$ and $R_Z(\theta)$ are Pauli X and Z rotation gates, respectively. 
The vertical line with black dots is a controlled-Z (CZ) gate.
}\label{fig:ansatz}
\end{figure}

Here, we demonstrate our benchmarking problem numerically \cite{git_program}.
We select ansatz $U(\vec{\theta})$ as Fig.~\ref{fig:ansatz} and feed $\ket{0}^{\otimes N}$ as input.
This type of ansatz is used in many of past researches \cite{Kubler2020adaptiveoptimizer, PhysRevResearch.2.043246, Nakanishi2019, koczor2020quantum, arrasmith2020operator, McArdle_2019, yamamoto2019natural, Stokes2020quantumnatural,Gu2021,Gard2020}, and therefore important to examine performances of various optimizers on them.
For each layers of blue~(light) and red~(dark) gates in Fig.~\ref{fig:ansatz}\subref{fig:blickwork}, we assign the same angles $\theta_i$ to all rotation gates (Fig.~\ref{fig:ansatz} \subref{subfig:b}).
We prepare an MPS representation of $U(\vec{\theta})\ket{0}^{\otimes N}$ following the method in Ref. \cite{Biamonte2017}. See Appendix A for the concrete implementation.
The MPS has bond dimension $D=2^3$.

The ansatz is translationally invariant, and hence its corresponding MPS representation and parent Hamiltonian also are.
This choice of ansatz has a nice property other than making the optimization problem simple, that is, we can guarantee that the problem Hamiltonian $H_{\mathrm{parent}}$ has a finite gap, which makes the optimization problem easier.
For a translationally invariant MPS, it is known that the corresponding parent Hamiltonian is gapped even at the thermodynamic limit if the MPS satisfies the so-called injectivity condition \cite{Perez-Garcia2007}.
We numerically checked that the ansatz in Fig.~\ref{fig:ansatz} is injective to ensure the gappedness.
An MPS represented by the set of matrices $\{A^{[k]}_{i_k}\}$ is injective iff the linear map $\Gamma_N: \mathcal{M}_D \rightarrow (\mathbb{C^d}^{\otimes L})$,
  \begin{equation}
    \Gamma_L (X) = \sum_{i_1, i_2, \cdots, i_L = 0}^{d-1} {\rm Tr}\left[ A_{i_1}, \cdots, A_{i_L} X \right]\ket{i_1, i_2, \cdots, i_L}
    \end{equation}
is injective for some integer $L$. 
Equivalently, an MPS is injective when the dimension of the space spanned by $\Gamma_L (X)$ is $D^2$ for some interger $L$.  
The smallest integer $L$ such that $\Gamma_L (X)$ becomes injective is called injectivity length of an MPS. 
From the numerical calculation, we find that injectivity length of our ansatz state is $7$ for most of the parameter choices.

\begin{figure}
\subfloat[\label{fig:syssize_locality}]{%
\includegraphics[width=0.75\linewidth]{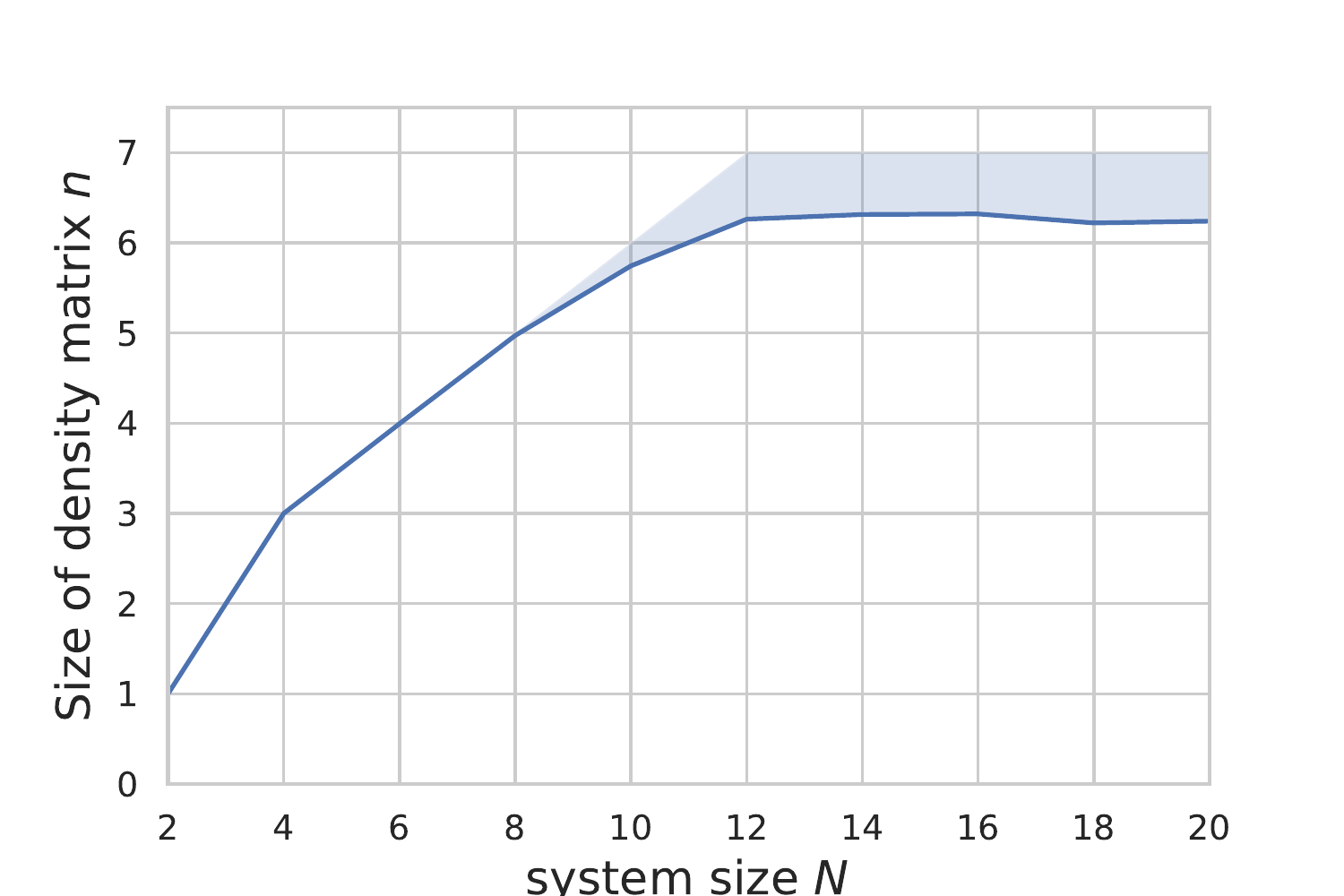}%
}\hfill%
\subfloat[\label{fig:Energyspectrum}]{%
  \includegraphics[width=0.8\linewidth]{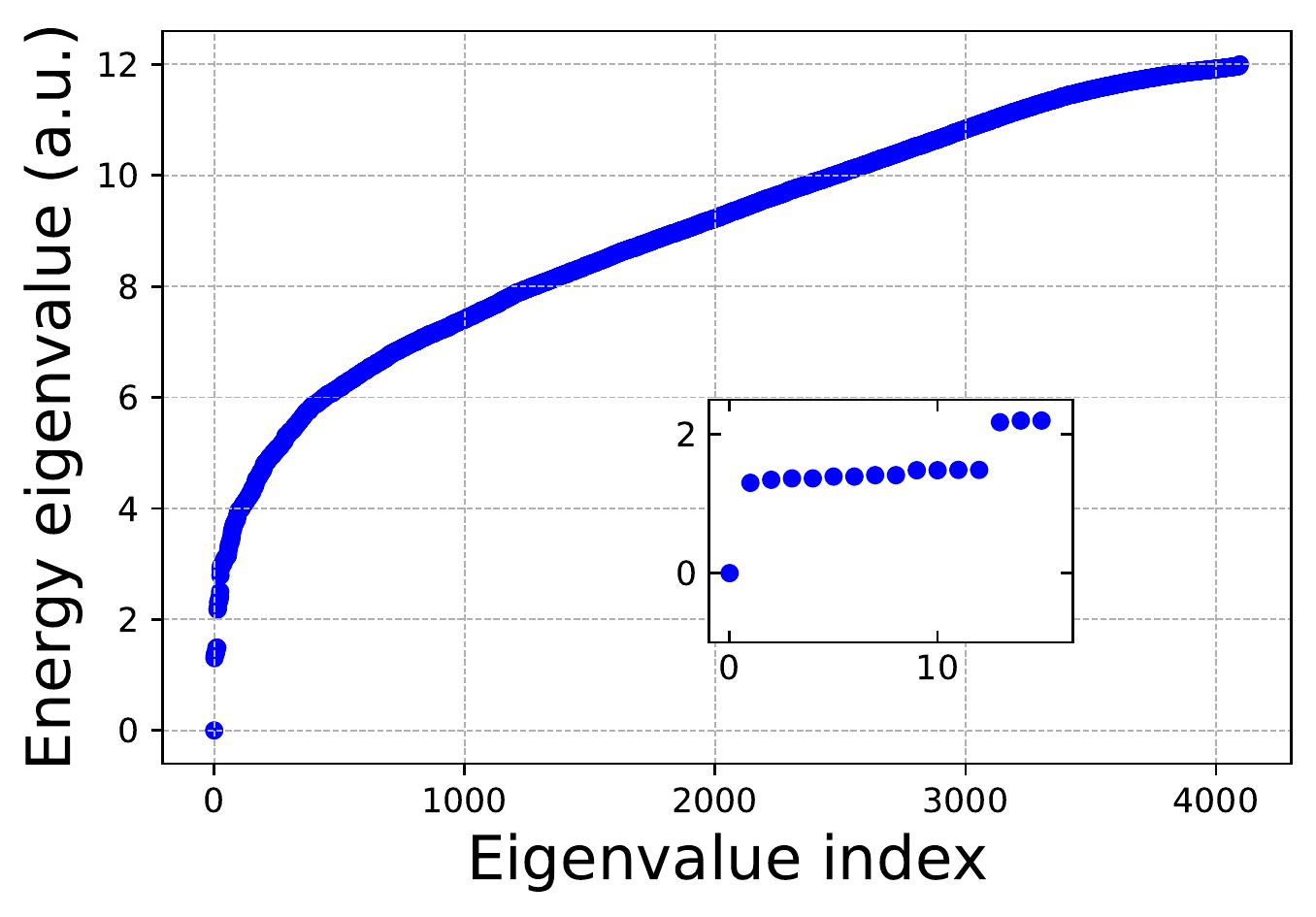}%
}%
\caption{
  (a) The smallest integer $n$ for which $\mathrm{Ker}(\rho^{(n)})$ becomes non-null for different system sizes. The bold line represents its average values. The upper bound of shaded area represents the maximum of such $n$ among 100 random choices of parameters.
  (b) The energy spectrum of our parent Hamiltonian at $N=12$ qubits. The inset is a zoom-in to the low-lying energies.
}\label{fig:injectivity}
\end{figure}


For step 1 of the algorithm, we chose $\vec{\theta}_{\mathrm{ans}}$ from uniform distribution on $[0,2\pi]$.
At step 2, we calculated $\mathrm{Ker}\,\rho^{(n)}$ starting from $n$ = 1.
Figure~\ref{fig:syssize_locality} shows the smallest integer $n$ for which $\mathrm{Ker}(\rho^{(n)})$ becomes non-null for different $N$.
It shows that taking $n=7$ or smaller is sufficient to make $\mathrm{Ker}\,\rho^{(n)}$ non-null.
This means that the parent Hamiltonian can be constructed with at most 7-body interactions whereas the trivial benchmarking Hamiltonian $U(\vec{\theta})\left(-\sum_i Z_i\right)U^{\dagger}(\vec{\theta})$ should generally have 12-body interactions as can be expected from the form of the ansatz in Fig. \ref{fig:ansatz}.
Thereby, we prove that the generated benchmarking problems are non-trivial.
Note that this integer $n$ such that $\mathrm{Ker}\,\rho^{(n)}$ becomes non-null generally depends on the depth of the ansatz, and the result $n=7$ is specific for our particular choice.
Finally, we generated the parent Hamiltonian by step 3.
We performed the exact diagonalization of the generated parent Hamiltonian to examine its properties.
Throughout this simulation, we set depth to be 3, counting a pair of blue~(light) and red~(dark) layers as depth 1 (see Fig. 1).
Each pair of blue~(light) and red~(dark) layers have 2 independent parameters.
Hence the number of independent parameters is 6.
The concrete form of the Hamiltonian is shown in Appendix B.
Its spectrum is shown in Fig. \ref{fig:Energyspectrum}.
We can observe that the ground state is unique and it has a certain energy gap.

We run numerical simulations of VQE on the parent Hamiltonian created by the above method.
We evaluate the performance of a typical gradient method, namely conjugate gradient (CG) method, and two typical quasi-Newton optimizers, namely the Broyden-Fletcher-Goldfarb-Shanno (BFGS) algorithm and sequential least squares programming (SLSQP) optimizer, for demonstration purposes.
The initial parameter $\vec{\theta}_{\mathrm{init}}$ of VQE is set by,
\begin{eqnarray}\label{eq:parameter}
\vec{\theta}_{\mathrm{init}} = \vec{\theta}_{\mathrm{ans}} + r \vec{\theta}_{\mathrm{random}},
\end{eqnarray}
where $\vec{\theta}_{\mathrm{random}}$ is a random parameter vector and its elements $\theta^{[i]}_{\mathrm{random}}$ are sampled from uniform distribution on the point on the unit sphere.
We vary $r$ from $0$ to $\pi$ for each optimizer to see how the distance between initial parameters and the solution affects the convergence of VQE. For each $r$, $100$ different $\vec{\theta}_{\mathrm{init}}$ are sampled.

\begin{figure}
\centering
\subfloat[\label{fig:defarence}]{%
  \includegraphics[width=0.9\linewidth]{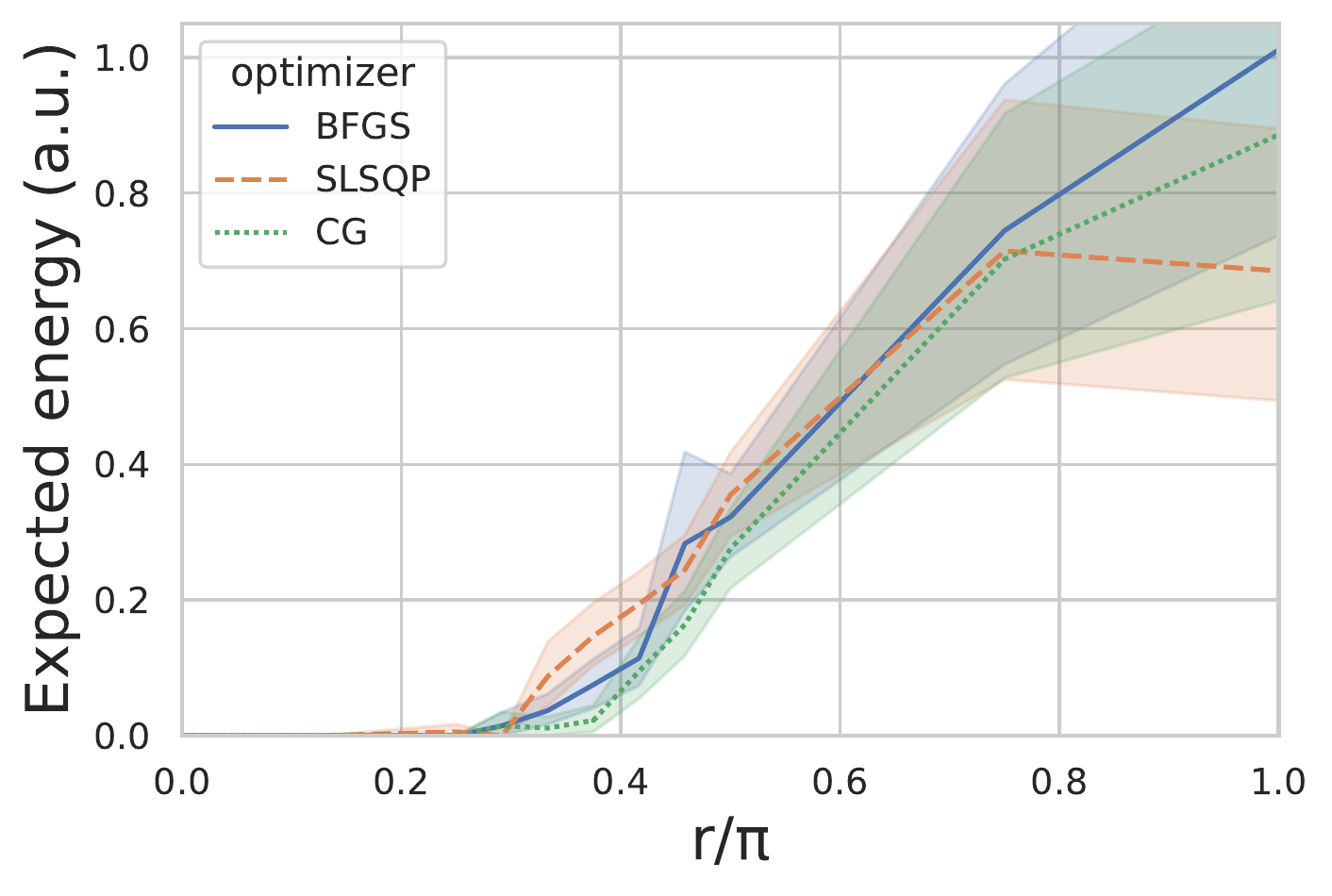}%
}\hfill
\centering
\subfloat[\label{fig:defarance-qubit}]{%
  \includegraphics[width=0.95\linewidth]{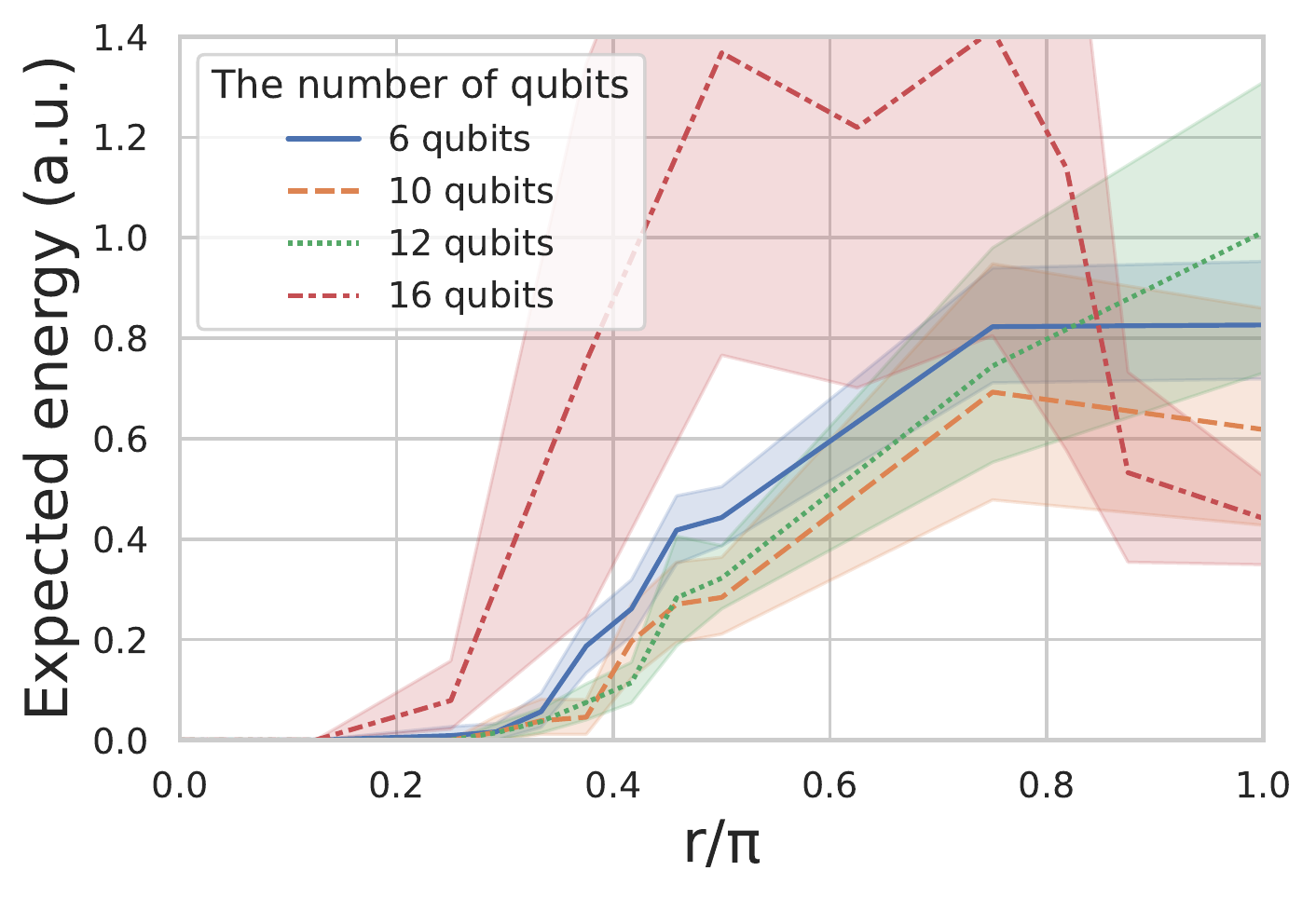}%
}

\caption{\label{fig:func_eval}
Optimized VQE energy with respect to the parameter $r$ in Eq. (\ref{eq:parameter}). The true ground-state energy is exactly 0. The line are the average value among 100 trials and the shaded area is the $90\%$ confidence interval. (a) Result using various optimizers. (b) Result using different system sizes $N$.
}
\end{figure}

Figure~\ref{fig:defarence} shows the converged energies of $N=12$-qubit systems using various optimizers.
Each optimizer exhibit similar performance.
Figure~\ref{fig:defarance-qubit} shows the converged energies using varying system sizes $N$ with BFGS optimizers.
We note that we used only 25 random initial guesses for $N=16$-qubit case to relax the computational resource required.
Notably, when $r \leq \pi/4$, they can achieve the exact ground state in most cases.
On the other hand, when $r \geq \pi/4$, they fail to do so.
Moreover, this behavior does not significantly change with different system sizes.
This result implies that the optimizers can successfully find the solution even with initial parameters which are rather far from the optimum.
It is possible that the failures witnessed for $r>\pi/4$ are due to barren plateaus \cite{mcclean2018barren} in the optimization landscape.
It should be interesting to explore how this threshold-like behavior arises among different ansatzes and optimizers in the future. 

\section{Conclusion and Discussion}
In this paper, we proposed a systematic method for generating benchmark problems for VQE from a given ansatz using the parent Hamiltonian.
Our approach can generate problems whose difficulty only depends on ansatz circuits, and thus successfully separates the two main factors, namely the form of ansatz and Hamiltonian, that determines the complexity of the optimization problem encountered in the VQE.
We have performed numerical calculations using a one-dimensional ansatz with periodic boundary condition and translational invariance.
The result has implied that the initial parameters can be rather far from an optimum in this case.

The following are several interesting directions to explore.
The first is to use a similar method for different types of ansatzes. For example, we can construct similar benchmarks for ones without translational invariance or periodic boundary condition.
The second is to use an optimizer other than BFGS or SLSQP, such as Adam \cite{Kingma2015} and SG-MCMC \cite{Chen2016}.
Third, while we employed only one-dimensional parameterized quantum circuits,
the proposed method is applicable to general low-depth parameterized quantum circuits 
including a low-depth two-dimensional quantum circuit corresponding to projected entangled pair state (PEPS) \cite{cirac2020matrix}.
The reduced density operators can be calculated by numerical simulation of quantum computation of relatively small size. Otherwise, we may employ an actual quantum device to estimate reduced density operators with an experiment, namely {\it hardware-efficient benchmark}, to construct a parent Hamiltonian.
Through these explorations, we might gain an implication for establishing a better optimizer for VQE.
Finally, it would be interesting to create a benchmark problem for other variational quantum algorithms (VQAs), such as variational linear system solver, using a similar method presented here.
We believe that the proposed benchmark construction can widely be used to analyze and improve the performance of VQAs.

\begin{acknowledgments}
FK would like to thank Ryotaro Suzuki and Tomohiro Yamazaki for useful discussions.
KM is supported by JST PRESTO Grant No. JPMJPR2019 and JSPS KAKENHI Grant No. 20K22330.
KF is supported by JSPS KAKENHI Grant No. 16H02211,  JST ERATO JPMJER1601, and JST CREST JPMJCR1673.
This work is supported by MEXT Quantum Leap Flagship Program (MEXT QLEAP) Grant Number JPMXS0118067394 and JPMXS0120319794, and laboratory rotation as part of the Program for Leading Graduate Schools: INTERACTIVE MATERIALS SCIENCE CADET in Osaka University.
We also acknowledge support from JST COI-NEXT program.
\end{acknowledgments}

\appendix
\section{Construction of MPS from an ansatz state}\label{app}
\begin{figure}[h]
    \centering
  \includegraphics[width=0.4\linewidth]{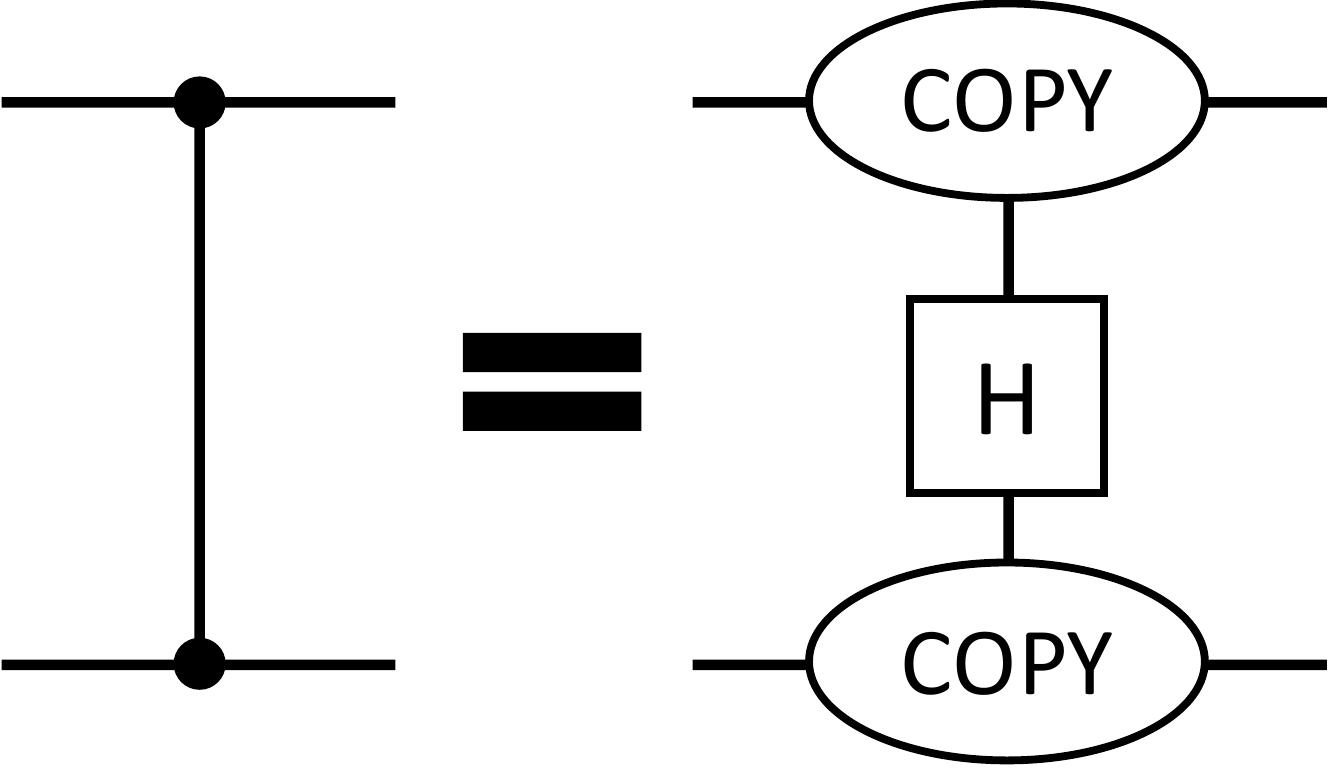}%
    \caption{Decomposition of a CZ gate into pairs of COPY tensor and Hadmord gate matrices as tensor network.
\label{fig:CZ}}
    \label{fig:my_label}
\end{figure}

Let COPY tensor be,
\begin{eqnarray}
    \mathrm{COPY}^{ij}_k = (1-i)(1-j)(1-k) + ijk,
\end{eqnarray}
and Hadamard tensor $H_i^j$ be,
\begin{equation}
H_i^j = (-1)^{ij},
\end{equation}
where $i,j,k \in\{0,1\}$.
Controlled-Z (CZ) gate in the circuit can be decomposed as,
\begin{eqnarray}
    \mathrm{CZ}^{qr}_{ij}   = \sum_{m,n} \mathrm{COPY}^{qm}_i H^n_{m} \mathrm{COPY}^r_{nj},
\end{eqnarray}
where $\mathrm{CZ}^{qr}_{ij} = \braket{qr|\mathrm{CZ}|ij}$.
Figure.~\ref{fig:CZ} shows the graphic representation of the decomposition.
The per site tensor of $A_i$ of the MPS can be obtained by calculating the contraction of the tensor in the row direction for each site.

\section{The parent Hamiltonian in our numerical experiment}\label{app1}

The local term $h_0$ of the parent Hamiltonian expanded in Pauli basis and its coefficients are shown in Table~\ref{table:coeff}.
In order to exclude the terms with very small coefficients that is computational error, we ignore the small values and only show ones with coefficients in the table.
Note that we can reconstruct the whole parent Hamiltonian from this information by considering its translational invariance.

\newpage
\begin{longtable}[h]{lcl}
\caption{Coefficients of local term $h_0$ in Pauli basis}
\label{table:coeff}\\
\hline
{} &  coefficient &  Pauli basis \\
\hline \hline
0   &     0.500000 &  $I_0$$I_1$$I_2$$I_3$$I_4$$I_5$$I_6$ \\
1   &     0.000209 &  $I_0$$I_1$$I_2$$I_3$$X_4$$X_5$$Z_6$ \\
2   &     0.000492 &  $I_0$$I_1$$I_2$$I_3$$Y_4$$Y_5$$Z_6$ \\
3   &     0.089243 &  $I_0$$I_1$$I_2$$X_3$$Y_4$$Z_5$$I_6$ \\
4   &     0.212358 &  $I_0$$I_1$$I_2$$Y_3$$I_4$$I_5$$I_6$ \\
5   &     0.045497 &  $I_0$$I_1$$I_2$$Y_3$$Y_4$$Z_5$$I_6$ \\
6   &     0.000295 &  $I_0$$I_1$$I_2$$Z_3$$I_4$$X_5$$Z_6$ \\
7   &     0.000324 &  $I_0$$I_1$$I_2$$Z_3$$I_4$$Y_5$$Z_6$ \\
8   &     0.007633 &  $I_0$$I_1$$X_2$$I_3$$X_4$$Z_5$$I_6$ \\
9   &     0.008401 &  $I_0$$I_1$$X_2$$I_3$$Y_4$$Z_5$$I_6$ \\
10  &     0.044993 &  $I_0$$I_1$$X_2$$I_3$$Z_4$$Z_5$$I_6$ \\
11  &     0.026537 &  $I_0$$I_1$$X_2$$X_3$$I_4$$Z_5$$I_6$ \\
12  &     0.004207 &  $I_0$$I_1$$X_2$$X_3$$X_4$$X_5$$Z_6$ \\
13  &     0.027735 &  $I_0$$I_1$$X_2$$X_3$$Y_4$$Y_5$$Z_6$ \\
14  &     0.006840 &  $I_0$$I_1$$X_2$$Y_3$$X_4$$X_5$$Z_6$ \\
15  &     0.029754 &  $I_0$$I_1$$X_2$$Y_3$$Y_4$$Y_5$$Z_6$ \\
16  &     0.011223 &  $I_0$$I_1$$X_2$$Y_3$$Z_4$$X_5$$Z_6$ \\
17  &     0.012353 &  $I_0$$I_1$$X_2$$Y_3$$Z_4$$Y_5$$Z_6$ \\
18  &     0.007544 &  $I_0$$I_1$$X_2$$Z_3$$X_4$$I_5$$I_6$ \\
19  &     0.037996 &  $I_0$$I_1$$X_2$$Z_3$$X_4$$Y_5$$Z_6$ \\
20  &     0.090413 &  $I_0$$I_1$$X_2$$Z_3$$Y_4$$I_5$$I_6$ \\
21  &     0.003490 &  $I_0$$I_1$$X_2$$Z_3$$Y_4$$X_5$$Z_6$ \\
22  &     0.009408 &  $I_0$$I_1$$Y_2$$I_3$$I_4$$X_5$$Z_6$ \\
23  &     0.010356 &  $I_0$$I_1$$Y_2$$I_3$$I_4$$Y_5$$Z_6$ \\
24  &     0.033226 &  $I_0$$I_1$$Y_2$$X_3$$X_4$$X_5$$Z_6$ \\
25  &     0.003822 &  $I_0$$I_1$$Y_2$$X_3$$Y_4$$Y_5$$Z_6$ \\
26  &     0.016866 &  $I_0$$I_1$$Y_2$$X_3$$Z_4$$I_5$$I_6$ \\
27  &     0.008900 &  $I_0$$I_1$$Y_2$$X_3$$Z_4$$X_5$$Z_6$ \\
28  &     0.009796 &  $I_0$$I_1$$Y_2$$X_3$$Z_4$$Y_5$$Z_6$ \\
29  &     0.025979 &  $I_0$$I_1$$Y_2$$Y_3$$I_4$$Z_5$$I_6$ \\
30  &     0.030839 &  $I_0$$I_1$$Y_2$$Y_3$$X_4$$X_5$$Z_6$ \\
31  &     0.006214 &  $I_0$$I_1$$Y_2$$Y_3$$Y_4$$Y_5$$Z_6$ \\
32  &     0.015778 &  $I_0$$I_1$$Y_2$$Y_3$$Z_4$$I_5$$I_6$ \\
33  &     0.090413 &  $I_0$$I_1$$Y_2$$Z_3$$X_4$$I_5$$I_6$ \\
34  &     0.060384 &  $I_0$$I_1$$Y_2$$Z_3$$X_4$$X_5$$Z_6$ \\
35  &     0.041823 &  $I_0$$I_1$$Y_2$$Z_3$$Y_4$$X_5$$Z_6$ \\
36  &     0.034325 &  $I_0$$I_1$$Y_2$$Z_3$$Z_4$$I_5$$I_6$ \\
37  &     0.008245 &  $I_0$$I_1$$Z_2$$I_3$$Z_4$$Z_5$$I_6$ \\
38  &     0.009443 &  $I_0$$I_1$$Z_2$$X_3$$I_4$$Z_5$$I_6$ \\
39  &     0.007088 &  $I_0$$I_1$$Z_2$$X_3$$X_4$$Y_5$$Z_6$ \\
40  &     0.016866 &  $I_0$$I_1$$Z_2$$X_3$$Y_4$$I_5$$I_6$ \\
41  &     0.002731 &  $I_0$$I_1$$Z_2$$X_3$$Y_4$$Y_5$$Z_6$ \\
42  &     0.018085 &  $I_0$$I_1$$Z_2$$X_3$$Z_4$$I_5$$I_6$ \\
43  &     0.006631 &  $I_0$$I_1$$Z_2$$Y_3$$X_4$$Y_5$$Z_6$ \\
44  &     0.015778 &  $I_0$$I_1$$Z_2$$Y_3$$Y_4$$I_5$$I_6$ \\
45  &     0.002555 &  $I_0$$I_1$$Z_2$$Y_3$$Y_4$$Y_5$$Z_6$ \\
46  &     0.020778 &  $I_0$$I_1$$Z_2$$Y_3$$Z_4$$I_5$$I_6$ \\
47  &     0.004242 &  $I_0$$I_1$$Z_2$$Y_3$$Z_4$$X_5$$Z_6$ \\
48  &     0.004669 &  $I_0$$I_1$$Z_2$$Y_3$$Z_4$$Y_5$$Z_6$ \\
49  &     0.014425 &  $I_0$$I_1$$Z_2$$Z_3$$X_4$$Y_5$$Z_6$ \\
50  &     0.034325 &  $I_0$$I_1$$Z_2$$Z_3$$Y_4$$I_5$$I_6$ \\
51  &     0.007633 &  $I_0$$I_1$$Z_2$$Z_3$$Y_4$$Y_5$$Z_6$ \\
52  &     0.044157 &  $I_0$$I_1$$Z_2$$Z_3$$Z_4$$I_5$$I_6$ \\
53  &     0.023271 &  $I_0$$Z_1$$I_2$$X_3$$I_4$$Z_5$$I_6$ \\
54  &     0.026537 &  $I_0$$Z_1$$I_2$$X_3$$X_4$$I_5$$I_6$ \\
55  &     0.009796 &  $I_0$$Z_1$$I_2$$X_3$$X_4$$X_5$$Z_6$ \\
56  &     0.012275 &  $I_0$$Z_1$$I_2$$X_3$$Y_4$$X_5$$Z_6$ \\
57  &     0.009443 &  $I_0$$Z_1$$I_2$$X_3$$Z_4$$I_5$$I_6$ \\
58  &     0.021175 &  $I_0$$Z_1$$I_2$$Y_3$$I_4$$Z_5$$I_6$ \\
59  &     0.010918 &  $I_0$$Z_1$$I_2$$Y_3$$X_4$$Y_5$$Z_6$ \\
60  &     0.025979 &  $I_0$$Z_1$$I_2$$Y_3$$Y_4$$I_5$$I_6$ \\
61  &     0.011223 &  $I_0$$Z_1$$I_2$$Y_3$$Y_4$$Y_5$$Z_6$ \\
62  &     0.007633 &  $I_0$$Z_1$$X_2$$I_3$$X_4$$I_5$$I_6$ \\
63  &     0.006673 &  $I_0$$Z_1$$X_2$$I_3$$X_4$$X_5$$Z_6$ \\
64  &     0.003531 &  $I_0$$Z_1$$X_2$$I_3$$Y_4$$X_5$$Z_6$ \\
65  &     0.023165 &  $I_0$$Z_1$$X_2$$X_3$$X_4$$Z_5$$I_6$ \\
66  &     0.041281 &  $I_0$$Z_1$$X_2$$Y_3$$Y_4$$Z_5$$I_6$ \\
67  &     0.001456 &  $I_0$$Z_1$$X_2$$Z_3$$I_4$$X_5$$Z_6$ \\
68  &     0.001603 &  $I_0$$Z_1$$X_2$$Z_3$$I_4$$Y_5$$Z_6$ \\
69  &     0.007799 &  $I_0$$Z_1$$X_2$$Z_3$$X_4$$Z_5$$I_6$ \\
70  &     0.008584 &  $I_0$$Z_1$$X_2$$Z_3$$Y_4$$Z_5$$I_6$ \\
71  &     0.008401 &  $I_0$$Z_1$$Y_2$$I_3$$X_4$$I_5$$I_6$ \\
72  &     0.007345 &  $I_0$$Z_1$$Y_2$$I_3$$X_4$$X_5$$Z_6$ \\
73  &     0.003886 &  $I_0$$Z_1$$Y_2$$I_3$$Y_4$$X_5$$Z_6$ \\
74  &     0.089243 &  $I_0$$Z_1$$Y_2$$X_3$$I_4$$I_5$$I_6$ \\
75  &     0.019120 &  $I_0$$Z_1$$Y_2$$X_3$$Y_4$$Z_5$$I_6$ \\
76  &     0.045497 &  $I_0$$Z_1$$Y_2$$Y_3$$I_4$$I_5$$I_6$ \\
77  &     0.041281 &  $I_0$$Z_1$$Y_2$$Y_3$$X_4$$Z_5$$I_6$ \\
78  &     0.001603 &  $I_0$$Z_1$$Y_2$$Z_3$$I_4$$X_5$$Z_6$ \\
79  &     0.001764 &  $I_0$$Z_1$$Y_2$$Z_3$$I_4$$Y_5$$Z_6$ \\
80  &     0.008584 &  $I_0$$Z_1$$Y_2$$Z_3$$X_4$$Z_5$$I_6$ \\
81  &     0.009449 &  $I_0$$Z_1$$Y_2$$Z_3$$Y_4$$Z_5$$I_6$ \\
82  &     0.044993 &  $I_0$$Z_1$$Z_2$$I_3$$X_4$$I_5$$I_6$ \\
83  &     0.010356 &  $I_0$$Z_1$$Z_2$$I_3$$X_4$$X_5$$Z_6$ \\
84  &     0.020813 &  $I_0$$Z_1$$Z_2$$I_3$$Y_4$$X_5$$Z_6$ \\
85  &     0.008245 &  $I_0$$Z_1$$Z_2$$I_3$$Z_4$$I_5$$I_6$ \\
86  &     0.008552 &  $I_0$$Z_1$$Z_2$$Z_3$$Z_4$$Z_5$$I_6$ \\
87  &     0.003954 &  $Z_0$$X_1$$I_2$$I_3$$X_4$$Y_5$$Z_6$ \\
88  &     0.009408 &  $Z_0$$X_1$$I_2$$I_3$$Y_4$$I_5$$I_6$ \\
89  &     0.007946 &  $Z_0$$X_1$$I_2$$I_3$$Y_4$$Y_5$$Z_6$ \\
90  &     0.000295 &  $Z_0$$X_1$$I_2$$Z_3$$I_4$$I_5$$I_6$ \\
91  &     0.001510 &  $Z_0$$X_1$$I_2$$Z_3$$I_4$$X_5$$Z_6$ \\
92  &     0.001663 &  $Z_0$$X_1$$I_2$$Z_3$$I_4$$Y_5$$Z_6$ \\
93  &     0.001456 &  $Z_0$$X_1$$I_2$$Z_3$$X_4$$Z_5$$I_6$ \\
94  &     0.001603 &  $Z_0$$X_1$$I_2$$Z_3$$Y_4$$Z_5$$I_6$ \\
95  &     0.000209 &  $Z_0$$X_1$$X_2$$I_3$$I_4$$I_5$$I_6$ \\
96  &     0.006673 &  $Z_0$$X_1$$X_2$$I_3$$X_4$$Z_5$$I_6$ \\
97  &     0.007345 &  $Z_0$$X_1$$X_2$$I_3$$Y_4$$Z_5$$I_6$ \\
98  &     0.010356 &  $Z_0$$X_1$$X_2$$I_3$$Z_4$$Z_5$$I_6$ \\
99  &     0.009796 &  $Z_0$$X_1$$X_2$$X_3$$I_4$$Z_5$$I_6$ \\
100 &     0.004207 &  $Z_0$$X_1$$X_2$$X_3$$X_4$$I_5$$I_6$ \\
101 &     0.013963 &  $Z_0$$X_1$$X_2$$X_3$$X_4$$Y_5$$Z_6$ \\
102 &     0.033226 &  $Z_0$$X_1$$X_2$$X_3$$Y_4$$I_5$$I_6$ \\
103 &     0.001946 &  $Z_0$$X_1$$X_2$$X_3$$Y_4$$X_5$$Z_6$ \\
104 &     0.006840 &  $Z_0$$X_1$$X_2$$Y_3$$X_4$$I_5$$I_6$ \\
105 &     0.012960 &  $Z_0$$X_1$$X_2$$Y_3$$X_4$$Y_5$$Z_6$ \\
106 &     0.030839 &  $Z_0$$X_1$$X_2$$Y_3$$Y_4$$I_5$$I_6$ \\
107 &     0.003164 &  $Z_0$$X_1$$X_2$$Y_3$$Y_4$$X_5$$Z_6$ \\
108 &     0.005050 &  $Z_0$$X_1$$X_2$$Y_3$$Z_4$$X_5$$Z_6$ \\
109 &     0.005559 &  $Z_0$$X_1$$X_2$$Y_3$$Z_4$$Y_5$$Z_6$ \\
110 &     0.025376 &  $Z_0$$X_1$$X_2$$Z_3$$X_4$$Y_5$$Z_6$ \\
111 &     0.060384 &  $Z_0$$X_1$$X_2$$Z_3$$Y_4$$I_5$$I_6$ \\
112 &     0.017576 &  $Z_0$$X_1$$X_2$$Z_3$$Y_4$$Y_5$$Z_6$ \\
113 &     0.003531 &  $Z_0$$X_1$$Y_2$$I_3$$X_4$$Z_5$$I_6$ \\
114 &     0.003886 &  $Z_0$$X_1$$Y_2$$I_3$$Y_4$$Z_5$$I_6$ \\
115 &     0.020813 &  $Z_0$$X_1$$Y_2$$I_3$$Z_4$$Z_5$$I_6$ \\
116 &     0.012275 &  $Z_0$$X_1$$Y_2$$X_3$$I_4$$Z_5$$I_6$ \\
117 &     0.001946 &  $Z_0$$X_1$$Y_2$$X_3$$X_4$$X_5$$Z_6$ \\
118 &     0.012829 &  $Z_0$$X_1$$Y_2$$X_3$$Y_4$$Y_5$$Z_6$ \\
119 &     0.003164 &  $Z_0$$X_1$$Y_2$$Y_3$$X_4$$X_5$$Z_6$ \\
120 &     0.013764 &  $Z_0$$X_1$$Y_2$$Y_3$$Y_4$$Y_5$$Z_6$ \\
121 &     0.005192 &  $Z_0$$X_1$$Y_2$$Y_3$$Z_4$$X_5$$Z_6$ \\
122 &     0.005714 &  $Z_0$$X_1$$Y_2$$Y_3$$Z_4$$Y_5$$Z_6$ \\
123 &     0.003490 &  $Z_0$$X_1$$Y_2$$Z_3$$X_4$$I_5$$I_6$ \\
124 &     0.017576 &  $Z_0$$X_1$$Y_2$$Z_3$$X_4$$Y_5$$Z_6$ \\
125 &     0.041823 &  $Z_0$$X_1$$Y_2$$Z_3$$Y_4$$I_5$$I_6$ \\
126 &     0.001614 &  $Z_0$$X_1$$Y_2$$Z_3$$Y_4$$X_5$$Z_6$ \\
127 &     0.003740 &  $Z_0$$X_1$$Z_2$$X_3$$X_4$$Y_5$$Z_6$ \\
128 &     0.008900 &  $Z_0$$X_1$$Z_2$$X_3$$Y_4$$I_5$$I_6$ \\
129 &     0.004687 &  $Z_0$$X_1$$Z_2$$X_3$$Y_4$$Y_5$$Z_6$ \\
130 &     0.004110 &  $Z_0$$X_1$$Z_2$$X_3$$Z_4$$X_5$$Z_6$ \\
131 &     0.004524 &  $Z_0$$X_1$$Z_2$$X_3$$Z_4$$Y_5$$Z_6$ \\
132 &     0.011223 &  $Z_0$$X_1$$Z_2$$Y_3$$X_4$$I_5$$I_6$ \\
133 &     0.005050 &  $Z_0$$X_1$$Z_2$$Y_3$$X_4$$X_5$$Z_6$ \\
134 &     0.005192 &  $Z_0$$X_1$$Z_2$$Y_3$$Y_4$$X_5$$Z_6$ \\
135 &     0.004242 &  $Z_0$$X_1$$Z_2$$Y_3$$Z_4$$I_5$$I_6$ \\
136 &     0.003740 &  $Z_0$$X_1$$Z_2$$Y_3$$Z_4$$X_5$$Z_6$ \\
137 &     0.004116 &  $Z_0$$X_1$$Z_2$$Y_3$$Z_4$$Y_5$$Z_6$ \\
138 &     0.004352 &  $Z_0$$Y_1$$I_2$$I_3$$X_4$$Y_5$$Z_6$ \\
139 &     0.010356 &  $Z_0$$Y_1$$I_2$$I_3$$Y_4$$I_5$$I_6$ \\
140 &     0.008746 &  $Z_0$$Y_1$$I_2$$I_3$$Y_4$$Y_5$$Z_6$ \\
141 &     0.000324 &  $Z_0$$Y_1$$I_2$$Z_3$$I_4$$I_5$$I_6$ \\
142 &     0.001663 &  $Z_0$$Y_1$$I_2$$Z_3$$I_4$$X_5$$Z_6$ \\
143 &     0.001830 &  $Z_0$$Y_1$$I_2$$Z_3$$I_4$$Y_5$$Z_6$ \\
144 &     0.001603 &  $Z_0$$Y_1$$I_2$$Z_3$$X_4$$Z_5$$I_6$ \\
145 &     0.001764 &  $Z_0$$Y_1$$I_2$$Z_3$$Y_4$$Z_5$$I_6$ \\
146 &     0.003954 &  $Z_0$$Y_1$$X_2$$I_3$$I_4$$X_5$$Z_6$ \\
147 &     0.004352 &  $Z_0$$Y_1$$X_2$$I_3$$I_4$$Y_5$$Z_6$ \\
148 &     0.013963 &  $Z_0$$Y_1$$X_2$$X_3$$X_4$$X_5$$Z_6$ \\
149 &     0.001606 &  $Z_0$$Y_1$$X_2$$X_3$$Y_4$$Y_5$$Z_6$ \\
150 &     0.007088 &  $Z_0$$Y_1$$X_2$$X_3$$Z_4$$I_5$$I_6$ \\
151 &     0.003740 &  $Z_0$$Y_1$$X_2$$X_3$$Z_4$$X_5$$Z_6$ \\
152 &     0.004117 &  $Z_0$$Y_1$$X_2$$X_3$$Z_4$$Y_5$$Z_6$ \\
153 &     0.010918 &  $Z_0$$Y_1$$X_2$$Y_3$$I_4$$Z_5$$I_6$ \\
154 &     0.012960 &  $Z_0$$Y_1$$X_2$$Y_3$$X_4$$X_5$$Z_6$ \\
155 &     0.002611 &  $Z_0$$Y_1$$X_2$$Y_3$$Y_4$$Y_5$$Z_6$ \\
156 &     0.006631 &  $Z_0$$Y_1$$X_2$$Y_3$$Z_4$$I_5$$I_6$ \\
157 &     0.037996 &  $Z_0$$Y_1$$X_2$$Z_3$$X_4$$I_5$$I_6$ \\
158 &     0.025376 &  $Z_0$$Y_1$$X_2$$Z_3$$X_4$$X_5$$Z_6$ \\
159 &     0.017576 &  $Z_0$$Y_1$$X_2$$Z_3$$Y_4$$X_5$$Z_6$ \\
160 &     0.014425 &  $Z_0$$Y_1$$X_2$$Z_3$$Z_4$$I_5$$I_6$ \\
161 &     0.000492 &  $Z_0$$Y_1$$Y_2$$I_3$$I_4$$I_5$$I_6$ \\
162 &     0.007946 &  $Z_0$$Y_1$$Y_2$$I_3$$I_4$$X_5$$Z_6$ \\
163 &     0.008746 &  $Z_0$$Y_1$$Y_2$$I_3$$I_4$$Y_5$$Z_6$ \\
164 &     0.027735 &  $Z_0$$Y_1$$Y_2$$X_3$$X_4$$I_5$$I_6$ \\
165 &     0.001606 &  $Z_0$$Y_1$$Y_2$$X_3$$X_4$$Y_5$$Z_6$ \\
166 &     0.003822 &  $Z_0$$Y_1$$Y_2$$X_3$$Y_4$$I_5$$I_6$ \\
167 &     0.012829 &  $Z_0$$Y_1$$Y_2$$X_3$$Y_4$$X_5$$Z_6$ \\
168 &     0.002731 &  $Z_0$$Y_1$$Y_2$$X_3$$Z_4$$I_5$$I_6$ \\
169 &     0.004687 &  $Z_0$$Y_1$$Y_2$$X_3$$Z_4$$X_5$$Z_6$ \\
170 &     0.005159 &  $Z_0$$Y_1$$Y_2$$X_3$$Z_4$$Y_5$$Z_6$ \\
171 &     0.011223 &  $Z_0$$Y_1$$Y_2$$Y_3$$I_4$$Z_5$$I_6$ \\
172 &     0.029754 &  $Z_0$$Y_1$$Y_2$$Y_3$$X_4$$I_5$$I_6$ \\
173 &     0.002611 &  $Z_0$$Y_1$$Y_2$$Y_3$$X_4$$Y_5$$Z_6$ \\
174 &     0.006214 &  $Z_0$$Y_1$$Y_2$$Y_3$$Y_4$$I_5$$I_6$ \\
175 &     0.013764 &  $Z_0$$Y_1$$Y_2$$Y_3$$Y_4$$X_5$$Z_6$ \\
176 &     0.002555 &  $Z_0$$Y_1$$Y_2$$Y_3$$Z_4$$I_5$$I_6$ \\
177 &     0.017576 &  $Z_0$$Y_1$$Y_2$$Z_3$$X_4$$X_5$$Z_6$ \\
178 &     0.001332 &  $Z_0$$Y_1$$Y_2$$Z_3$$Y_4$$Y_5$$Z_6$ \\
179 &     0.007633 &  $Z_0$$Y_1$$Y_2$$Z_3$$Z_4$$I_5$$I_6$ \\
180 &     0.004117 &  $Z_0$$Y_1$$Z_2$$X_3$$X_4$$Y_5$$Z_6$ \\
181 &     0.009796 &  $Z_0$$Y_1$$Z_2$$X_3$$Y_4$$I_5$$I_6$ \\
182 &     0.005159 &  $Z_0$$Y_1$$Z_2$$X_3$$Y_4$$Y_5$$Z_6$ \\
183 &     0.004524 &  $Z_0$$Y_1$$Z_2$$X_3$$Z_4$$X_5$$Z_6$ \\
184 &     0.004979 &  $Z_0$$Y_1$$Z_2$$X_3$$Z_4$$Y_5$$Z_6$ \\
185 &     0.012353 &  $Z_0$$Y_1$$Z_2$$Y_3$$X_4$$I_5$$I_6$ \\
186 &     0.005559 &  $Z_0$$Y_1$$Z_2$$Y_3$$X_4$$X_5$$Z_6$ \\
187 &     0.005714 &  $Z_0$$Y_1$$Z_2$$Y_3$$Y_4$$X_5$$Z_6$ \\
188 &     0.004669 &  $Z_0$$Y_1$$Z_2$$Y_3$$Z_4$$I_5$$I_6$ \\
189 &     0.004116 &  $Z_0$$Y_1$$Z_2$$Y_3$$Z_4$$X_5$$Z_6$ \\
190 &     0.004531 &  $Z_0$$Y_1$$Z_2$$Y_3$$Z_4$$Y_5$$Z_6$ \\
\hline
\end{longtable}


%

\end{document}